\documentclass[journal]{IEEEtran}
\usepackage{cite}
\usepackage{amsmath,amssymb,amsfonts}
\usepackage{algorithmic}
\usepackage{graphicx}
\usepackage{textcomp}
\usepackage{xcolor}
\usepackage[utf8]{inputenc}
\usepackage{amsmath}
\usepackage[english]{babel}
\usepackage{amsthm}
\usepackage{mathtools}
\usepackage{multirow}
\usepackage{mathtools}
\usepackage[ruled,norelsize]{algorithm2e}

\theoremstyle{definition}

\DeclarePairedDelimiter\floor{\lfloor}{\rfloor}
\newtheorem{prop}{Proposition}
\def\BibTeX{{\rm B\kern-.05em{\sc i\kern-.025em b}\kern-.08em
    T\kern-.1667em\lower.7ex\hbox{E}\kern-.125emX}}
 \newtheorem*{remark}{Remark}

\begin{document}

\bstctlcite{IEEEexample:BSTcontrol}

\title{$Q$-ary Multi-Mode OFDM with Index Modulation}

\author{Ferhat Yarkin~\IEEEmembership{Student~Member,~IEEE} and Justin P.~Coon~\IEEEmembership{Senior~Member,~IEEE}
\thanks{F. Yarkin and J. P. Coon are with the Department of Engineering Science, University of Oxford, Parks Road, Oxford, OX1 3PJ, U.K. E-mail: \{ferhat.yarkin and justin.coon\}@eng.ox.ac.uk}
}


%


\maketitle

\begin{abstract}
In this paper, we propose a novel orthogonal frequency division multiplexing with index modulation (OFDM-IM) scheme, which we call $Q$-ary multi-mode OFDM-IM ($Q$-MM-OFDM-IM). In the proposed scheme, $Q$ disjoint $M$-ary constellations are used repeatedly on each subcarrier, and a maximum-distance separable code is applied to the indices of these constellations to achieve the highest number of index symbols. A low-complexity subcarrier-wise detection is shown possible for the proposed scheme. Spectral efficiency (SE) and the error rate performance of the proposed scheme are further analyzed. It is shown that the proposed scheme exhibits a very flexible structure that is capable of encompassing conventional OFDM as a special case. It is also shown that the proposed scheme is capable of considerably outperforming the other OFDM-IM schemes and conventional OFDM in terms of error and SE performance while preserving a low-complexity structure.            
\end{abstract}


\IEEEpeerreviewmaketitle
\section{Introduction}

The studies, so far, show that index modulation (IM) techniques exhibit important advantages compared to conventional modulation techniques. Specifically, when IM is applied to orthogonal frequency division multiplexing (OFDM) technique, a better error performance and improved data rate compared to conventional OFDM are shown to be possible \cite{Basar2013,Fan2015, Mao2017, Mao20173, Wen2017, Wen2018}. Moreover, an application of a recent IM scheme, called set partition modulation (SPM), to OFDM brings about a marginal enhancement in error rate at high signal-to-noise ratio (SNR) values and a substantial improvement in data rate when compared to other OFDM-IM benchmarks \cite{Yarkin2019}. 

It is well-known that, for a fading channel, the error performance of a codebook is limited by the minimum Euclidean distance between codeword pairs that have the minimum Hamming distance \cite{Tarokh1998}. Moreover, such minimum Euclidean and Hamming distances, respectively, determine the coding and diversity gains of error probability curves related to the codebook. Hence, the superior error performance provided by the OFDM-IM schemes is based on the fact that the information bits are mapped not only to signal space as in conventional OFDM but also to the index domain \footnote{ Note that the minimum Hamming distance between conventional modulation symbols drawn from the signal space is limited to one; however, such distance is two for index symbols drawn from the index domain. Since the overall OFDM-IM symbols are formed by both conventional modulation and index symbols, the minimum Hamming distance between these symbols is limited to one.}. Such  a mapping enables the OFDM-IM schemes to have a higher minimum Euclidean distance in the signal space than conventional OFDM. In this regard, conventional OFDM-IM \cite{Basar2013, Fan2015} encodes data into the combinations of active subcarriers, thus the total power is distributed across all subcarriers. Moreover, dual-mode OFDM-IM (DM-OFDM-IM) \cite{Mao2017,Mao20173}, multi-mode OFDM-IM (MM-OFDM-IM) \cite{Wen2017,Wen2018} and OFDM-SPM \cite{Yarkin2019} use disjoint constellations on all subcarriers. The presence of the index codewords makes these schemes capable of achieving the same spectral efficiency (SE) as that of conventional OFDM by employing lower order modulation. On the other hand, the potential of index domain is exploited more by the studies that increase the number of index symbols with combinatorial tools such as permutation and set partitioning \cite{Wen2017,Wen2018,Yarkin2019}. However, it is not obvious whether the existing OFDM-IM schemes can achieve the full potential of the index domain by producing the largest number of possible index symbols.    

Against this background, we propose a novel MM-OFDM-IM scheme named $Q$-ary  MM-OFDM-IM ($Q$-MM-OFDM-IM) which is capable of exploiting the full potential of the index domain by employing $Q$ disjoint constellations repeatedly on each subcarrier.  In this scheme, unlike the MM-OFDM-IM scheme in \cite{Wen2017} that uses the permutations of disjoint constellations to form the index symbols, we employ a completely different approach that uses a maximum-distance separable (MDS) code on the disjoint constellations to achieve the highest number of index symbols. Then, for the proposed scheme, we present a sub-optimal maximum likelihood (ML) detector.  We also investigate the SE and bit error rate (BER) of $Q$-MM-OFDM-IM in this paper and obtain an upper-bound on the BER. Our analytical, as well as numerical findings, show that the proposed scheme can achieve a substantially better performance than OFDM-SPM, MM-OFDM-IM, OFDM-IM and conventional OFDM in terms of SE and BER while exhibiting a very simple structure and high flexibility.


 \section{System Model}\label{section:section2}

In this section, we present the system model of the $Q$-MM-OFDM-IM scheme.

\subsection{Transmitter} 

$m$ input bits enter the transmitter, and these bits are divided into $B=m/f$ blocks, each having $f$ input bits. Similarly, the total number of subcarriers $N_T$ is also divided into $B=N_T/N$ blocks, each having $N$ subcarriers. 

Since each bit and each subcarrier block has the same mapping operation, we focus on a single block, the $b$th block (where $b\in \big\{1, 2, \ldots, B\big\}$), in what follows. In the $b$th block, the $f$ information bits are further divided   into two parts, one of them having $f_1$ bits and the other one having $f_2$ bits with $f_1+f_2=f$.  Letting $Q$ be a positive integer, the first $f_1$ bits are used to determine the one of $Q$ disjoint $M$-ary constellations\footnote{For two disjoint constellations $\mathcal{M}_q$ and $\mathcal{M}_{\hat{q}}$ where $q,\hat{q}\in \big\{0, 1, \ldots, Q-1\big\}$ and $q \neq \hat{q}$, $\mathcal{M}_q \cap \mathcal{M}_{\hat{q}}=\emptyset$. Note also that we choose the size of each constellation as $M$, i.e., $|\mathcal{M}_q|=M, \forall q\in \big\{0, 1, \ldots, Q-1\big\}$, for convenience.} $\mathcal{M}_q$, $q\in\big\{0,1,\ldots, Q-1\big\}$, or in other words modes,  which will be used on each subcarrier. \emph{Unlike the other OFDM-IM schemes, which employ different modes on their subcarriers, the proposed scheme is capable of using any mode on any subcarrier, repeatedly.} Hence, we have $Q^N$ mode patterns in total on $N$ subcarriers.  

Unlike the index symbols of the OFDM-IM schemes, the minimum Hamming distance between  $Q^N$ patterns is limited to one. In this context, it is important to attain the highest number of the index symbols that achieve the same minimum Hamming distance as those of the OFDM-IM schemes. Such a number is bounded by the Singleton bound\footnote{ Note that each index pattern of the $Q$-MM-OFDM-IM scheme can be regarded as a $Q$-ary block code of length $N$ since the index bits are mapped to $N$-tuples whose elements are chosen among $Q$ disjoint constellations. Hence, the Singleton bound is valid for the number of index symbols of the $Q$-MM-OFDM-IM scheme.} \cite{Singleton64}. With $B_Q(d, N)$ and $d$ denoting the maximum number of possible codewords in a $Q$-ary block code of length $N$ and minimum Hamming distance $d$ between such codewords, the Singleton bound states that
\begin{align}
    B_Q(d,N)\leq Q^{N-d+1}.
\end{align}
To have the same minimum Hamming distance between the index symbols as the index symbols of the OFDM-IM schemes, one needs to pick $d=2$. In that case, the Singleton bound becomes $B_Q(2,N)\leq Q^{N-1}$. This bound can be achieved by a simple maximum-distance separable (MDS) code \cite{Singleton64}. Such a code forms the first $N-1$ elements, $I_{\tau},\tau\in \big\{1, \ldots, N-1\big\}$, by using the integers, $0,1,\ldots,Q-1$ as symbols, i.e., $I_{\tau}\in \big\{0, 1, \ldots, Q-1\big\}$, and the last symbol, $I_N$ is chosen from the same integers by letting the code be those $N$-tuples summing to zero under modulo-$Q$ arithmetic, i.e., $(I_1+I_2+\ldots+I_N) \bmod Q=0 $. Hence, there are $Q^{N-1}$ such $N$-tuples. Then, we map these $N$-tuples to the one of the index sets in the $b$th block, $\mathcal{I}^b\coloneqq\big\{I_{1}, I_{2}, \ldots, I_{N}\big\}$, and the $n$th element of the index set is used to determine the index of the $M$-ary constellation on the $n$th subcarrier where $n\in \big\{1, 2, \ldots, N\big\}$. $f_1$ bits are further mapped to one of these index sets, and therefore $f_1=\floor{\log_2 Q^{N-1}}$. Note that these operations result in $Q^{N-1}$ index symbols whose minimum Hamming distance is two.  Moreover, the mapping of $f_1$ bits to the index symbols can be implemented by using a look-up table. 

Let us consider an example of how we form the index symbols and implement the look-up table between these codewords and the information bits for the proposed scheme when $Q=N=3$. We use the integers, $0, 1, 2$, to form MDS codes as shown in the leftmost column of Table \ref{table:table1}. The corresponding triple codes are given in the second column (from the left) of the table. By mapping them to the index symbol vectors, we construct the final index sets of the $Q$-MM-OFDM-IM scheme as shown in the table. Moreover, $f_1$ bits are used to determine the specific index set.

 \begin{table}[t]
\centering
\caption{Index Symbol Generation and Look-Up Table Example for the Q-MM-OFDM-GSPM scheme when $Q=N=3$.}
\label{table:table1}
\begin{tabular}{|c|c|c|c|}
\hline
\begin{tabular}[c]{@{}c@{}}$(N-1)$-tuple\\ $(I_1, I_2)$\end{tabular} & \begin{tabular}[c]{@{}c@{}}$N$-tuple\\ MDS Code\end{tabular} & \begin{tabular}[c]{@{}c@{}}$Q$-MM-OFDM-IM\\ Index Set\end{tabular} & $f_1$ bits      \\ \hline
$(0, 0)$ & $(0, 0, 0)$ & $\big\{0, 0, 0\big\}$ & $[0 \: 0 \: 0]$ \\ \hline
$(0, 1)$ & $(0, 1, 2)$ & $\big\{0, 1, 2\big\}$ & $[0 \: 0 \: 1]$ \\ \hline
$(0, 2)$ & $(0, 2, 1)$ & $\big\{0, 2, 1\big\}$ & $[0 \: 1 \: 0]$ \\ \hline
$(1, 0)$ & $(1, 0, 2)$ & $\big\{1, 0, 2\big\}$ & $[0 \: 1 \: 1]$ \\ \hline
$(1, 1)$ & $(1, 1, 1)$ & $\big\{1, 1, 1\big\}$ & $[1 \: 0 \: 0]$ \\ \hline
$(1, 2)$ & $(1, 2, 0)$ & $\big\{1, 2, 0\big\}$ & $[1 \: 0 \: 1]$ \\ \hline
$(2, 0)$ & $(2, 0, 1)$ & $\big\{2, 0, 1\big\}$ & $[1 \: 1 \: 0]$ \\ \hline
$(2, 1)$ & $(2, 1, 0)$ & $\big\{2, 1, 0\big\}$ & $[1 \: 1 \: 1]$ \\ \hline
$(2, 2)$ & $(2, 2, 2)$ & $\big\{2, 2, 2\big\}$ & unused          \\ \hline
\end{tabular}
\end{table}
 
 Once the index set, $\mathcal{I}^b$, is determined by $f_1$ bits, the remaining $f_2$ bits are used to modulate symbols on the $N$ subcarriers by using the disjoint $M$-ary constellations\footnote{By following the useful design guidelines in \cite{Wen2017}, we obtain the disjoint PSK constellations $\mathcal{M}_q$ by rotating each constellation with the angle of $2q\pi/(M Q)$, $q=0, \ldots,N-1$, to maximize the distance between constellation points. To obtain disjoint QAM constellations, likewise \cite{Wen2017}, we employ the well-known set partitioning technique in \cite{Ungerboeck1982}.} regarding the determined index set.  Hence, the $Q$-MM-OFDM-IM symbol vector corresponding to the $b$th block can be written as $\textbf{s}^b=[s_{1}^b, s_{2}^b \ldots, s_{N}^b]$ where $s_{n}^b \in \mathcal{M}_{I_{n}} $, $I_n\in \big\{0, 1, \ldots, Q-1\big\}$ is the $n$th element of the set $\mathcal{I}^b$. Since $|\mathcal{M}_q|=M, \; \forall q \in \big\{0, 1, \ldots, Q-1\big\}$, $f_2=N\log_2 M$. After obtaining symbol vectors for all blocks, an OFDM block creator forms the overall symbol vector  $\textbf{s}\coloneqq [s(1), s(2), \ldots,s(N_T)]^T=[\textbf{s}^1,\ldots, \textbf{s}^b, \ldots, \textbf{s}^B]^T\in \mathcal{C}^{N_T\times 1}$. After this point, exactly the same operations as conventional OFDM are applied\footnote{We assume that the elements of $\textbf{s}$ are interleaved sufficiently and the maximum spacing is achieved for the subcarriers. It is also assumed that each modulated symbol carried by a subcarrier has unit energy, i.e., $\operatorname{E}[|s(t)|^2]=1$, $t=1,\ldots,N_T$.}.  
 
 \subsection{Receiver}
At the receiver, the received signal is down-converted, and the cyclic prefix is then removed from each received baseband symbol vector before processing with an FFT. After employing a $N_T$-point FFT operation, the frequency-domain received signal vector can be written as 
\begin{align}
    \textbf{y} \coloneqq [y(1),y(2), \ldots, y(N_T)]^T=\sqrt{E_S}\textbf{S}\textbf{h}+\textbf{n}
\end{align}
where $E_S$ is the energy of the transmitted symbol vector and $\textbf{S}=\text{diag}(\textbf{s})$. Moreover, $\textbf{h}$ and $\textbf{n}$ are $N_T\times 1$ channel and noise vectors, respectively. Elements of these vectors follow the complex-valued Gaussian distributions $\mathcal{CN}(0,1)$ and $\mathcal{CN}(0,N_0)$, respectively, where $N_0$ is the noise variance.

Since the encoding procedure for each block is independent of others, decoding can be performed independently at the receiver. Hence, using maximum likelihood (ML) detection, the detected symbol vector for the $b$th block can be written as 
\begin{align}\label{eq:eq2}
    ({\hat{\mathcal{I}}^b,\hat{\textbf{s}}^b})= \arg \min_{\mathcal{I}^b, \textbf{s}^b} ||\textbf{y}^b-\sqrt{E_S}\textbf{S}^b\textbf{h}^b||^2
\end{align}
where $\textbf{y}^b=[y((b-1)N+1), \ldots, y(bN)]^T$, $\textbf{S}^b=\text{diag}(\textbf{s}^b)$ and $\textbf{h}^b=[h((b-1)N+1), \ldots, h(bN)]^T$.

Optimum ML detection complexity of the proposed scheme is of order $ O(Q^{N-1}M^N)$ since we have $Q^{N-1}$ and $M^N$ index and $M$-ary modulation symbols, respectively, on $N$ subcarriers. Hence, such a detection mechanism is impractical when $N$ and $Q$ are high. To overcome the high complexity of the optimum ML detector, we design a low-complexity suboptimal ML detector which operates on $N-1$ subcarriers independently while the disjoint constellation on the remaining subcarrier is decided according to the disjoint constellations on these $N-1$ subcarriers based on the MDS code. The proposed low-complexity ML (LC-ML) detection is based on the fact that each disjoint constellation can be used on each subcarrier repeatedly, and the MDS code forms the index symbols in a way that the sum of elements of corresponding $N$-tuples is zero under modulo-$Q$ arithmetic. The proposed LC-ML detector for a $Q$-MM-OFDM-IM block\footnote{Here, we omit the block superscript for convenience since the decoding can be performed independently for each block.} can be explained as follows: 
\begin{enumerate}
    \item Sort the channel gains of $N$ subcarriers in descending order. In other words, $|h(\lambda_1)|^2\ge \ldots \ge |h(\lambda_N)|^2$ where $\lambda_n\in\big\{1,\ldots,N\big\}$.
    \item Determine the constellation, $\mathcal{M}_{I_{\lambda}}$, $\lambda \in \big\{\lambda_1, \ldots, \lambda_{N-1}\big\}$ and the modulation symbol, $s_{\lambda} \in \mathcal{M}_{I_{\lambda}}$, on each subcarrier by substituting all possible constellation symbols, which are drawn from the union of all disjoint constellations, $\mathcal{M}=\mathcal{M}_0\cup\ldots\cup\mathcal{M}_{Q-1}$, based on an ML detector except for the subcarrier that has the smallest channel gain, i.e., $\lambda_N$th subcarrier. The detected disjoint constellation index and the modulation symbol for the $\lambda$th subcarrier of an $Q$-MM-OFDM-IM block can be written as 
    \begin{align}
        \big(\hat{I}_{\lambda},~ \hat{s}_{\lambda}\big)=\arg \min_{I_{\lambda},~ s_{\lambda}}
        |y(\lambda)-h(\lambda)s_{\lambda}|^2.
    \end{align}
    \item Estimate the constellation index, $\hat{I}_{\lambda_N}$, on the $\lambda_N$th according to the equation 
    \begin{align}
       (\hat{I}_{\lambda_1}+\hat{I}_{\lambda_2}+\ldots+\hat{I}_{\lambda_N}) \bmod Q=0. 
    \end{align}
    \item Determine the modulation symbol, $\hat{s}_{\lambda_N}$, on the $\lambda_N$th subcarrier by substituting $M$ symbols belonging to the constellation $\mathcal{M}_{\hat{I}_{\lambda_N}}$. 
\end{enumerate}

The proposed LC-ML detector compares $QM(N-1)$ squared Euclidean distances for $N-1$ subcarriers, which have the highest $N-1$ channel gains, and $M$ squared Euclidean distances for the subcarrier that has the smallest channel gain. The total number of squared Euclidean distance comparisons for a $Q$-MM-OFDM-IM subcarrier is given by $QM-QM/N+M/N$. Hence, the computational complexity of the proposed detector is of order $O(QM)$. It can be shown that the proposed ML detector exhibits a lower complexity than the subcarrier-wise log-likelihood detector of MM-OFDM-IM \cite{Wen2017} when $Q=N$ and the constellation size of MM-OFDM-IM is chosen in a way that the overall SEs of both schemes are equal.

\section{Performance Analysis}\label{section:section4}
In this section, we analyze the SE and BER of the proposed scheme. 

\subsection{Spectral Efficiency}

As defined above, the proposed scheme transmits $f_1=\floor {\log_2 Q^{N-1}}$ and $f_2=N \log_2 M$  bits by the index symbols and the $M$-ary constellation symbols on $N$ subcarriers, respectively. Hence, by ignoring cyclic prefix length, the SE of the proposed scheme can be given by 
\begin{align}
    \eta=\frac{f_1+f_2}{N}=\frac{\floor {\log_2Q^{N-1}} +N \log_2 M}{N}
\end{align}
This SE scales as $\eta \sim \log_2(QM)$.

\begin{remark}
The proposed scheme exhibits outstanding flexibility since $Q$, $N$ and $M$ can be adjusted independently to achieve a desired SE for the proposed scheme. On the other hand, it is easy to notice that the proposed scheme is equivalent to conventional OFDM when $Q=1$. Moreover, when $Q=N$, we arrive at $N^{N-1}$ index codewords for the proposed scheme and the SE  scales as $\eta \sim \log_2 (NM)$, which is the highest SE among the OFDM-IM schemes having $N$ disjoint constellations and subcarriers. 
\end{remark}

\begin{prop}\label{prop:proposition1}
As $N\to\infty$, the index codewords of $Q$-MM-OFDM-IM are capable of achieving the same SE as the overall SEs of the MM-OFDM-IM and OFDM - Ordered Full SPM (OFDM-OFSPM) schemes, which employ $N$ disjoint $M$-ary constellations, when $Q=NM/e$ and $Q=NM/(e\ln2)$, respectively. 
\end{prop}

\begin{IEEEproof}
The proof can easily be obtained by looking at the SEs of MM-OFDM-IM and OFDM-OFSPM, and the SE provided by only the index symbols of the $Q$-MM-OFDM-IM scheme. They can be given, respectively, as $\eta \sim \log_2 N - \log_2 e + \log_2 M$ \cite{Wen2017}, $\eta \sim \log_2 N - \log_2 (e/\ln2) + \log_2 M$  \cite{Yarkin2019} and $\eta \sim \log_2 Q$ when $N\to\infty$.
\end{IEEEproof}

Based on the fact in Proposition \ref{prop:proposition1}, the proposed scheme can achieve the same SE as that of the MM-OFDM-IM scheme without employing conventional modulation while utilizing fewer constellation points. Such a result makes the proposed scheme not only spectrally more efficient but also more reliable in terms of error rate than MM-OFDM-IM since the index symbols are capable of achieving higher minimum Hamming distance than the conventional modulation symbols.

\subsection{Bit-Error Rate}

An upper-bound on the average BER is given by the well-known union bound
\begin{align}\label{eq:eq13}
    P_b \leq \frac{1}{f2^f}\sum_{i=1}^{2^f}\sum_{j=1}^{2^f}P(\textbf{S}^i\to\textbf{S}^j)D(\textbf{S}^i\to\textbf{S}^j)
\end{align}
where $P(\textbf{S}^i\to\textbf{S}^j)$ stands for the pairwise error probability (PEP) regarding the erroneous detection of $\textbf{S}^i$ as $\textbf{S}^j$ where $i \neq j$, $i,j\in \big\{1, \ldots,Q^{N-1}M^N\big\}$, $\textbf{S}^i=\text{diag}(\textbf{s}^i)$ and $\textbf{S}^j=\text{diag}(\textbf{s}^j)$ and $D(\textbf{S}^i\to\textbf{S}^j)$ is the number of bits in error for the corresponding pairwise error event. One can use the same PEP expression as in \cite{Basar2013} and substitute the $Q$-MM-OFDM-IM codewords to obtain the upper bound on the average BER.

As explained in the previous section, the minimum Hamming distance between index symbols of the $Q$-MM-OFDM-IM is equal to two. However, the minimum Hamming distance between the conventional modulation symbols is limited to one. Hence, as in other OFDM-IM schemes, the average BER expression of the proposed scheme will be dominated by the minimum Euclidean distance between the modulation symbols at high SNR values, and therefore, the diversity order of the BER curves is limited to one. However, the proposed scheme is capable of producing a large number of index symbols since the number of these symbols is a power of $Q$. Hence, one can use only the index symbols of the proposed scheme to attain a codebook. In this case, the diversity order of the BER curves becomes equal to two. 

\section{Numerical Results}\label{section:section5}

 In this section, we compare uncoded and coded BER performance of the proposed scheme with that of the OFDM-IM benchmarks and conventional OFDM. For coded schemes, we use a rate-1/3 turbo code that is specified by Third-Generation Partnership Project (3GPP) \cite{schlegel2004}. We also show the effectiveness of the proposed LC-ML detector compared to the optimum ML detector.

In figures, ``$Q$-MM-OFDM-IM $(Q, N, M)$'' and ``$Q$-MM-OFDM-IM $(Q, N, M)$, QAM'' signify the proposed schemes employing $Q$ disjoint $M$-ary PSK and QAM constellations, respectively, on $N$ subcarriers, whereas ``OFDM-OFSPM $(N, M)$'' is the variant of OFDM-SPM \cite{Yarkin2019} employing all set partitions and having $N$ subcarriers as well as $N$ disjoint $M$-PSK constellations in each OFDM block. ``OFDM-IM $(N, K_a, M)$'' stands for the conventional OFDM-IM scheme in which $K_a$ out of $N$ subcarriers are activated to send $M$-PSK modulated symbols in each block. Finally,  ``MM-OFDM-IM $(N, M)$'' represents a multi-mode scheme having $N$ subcarriers along with $N$ disjoint $M$-PSK constellations in each block.    Note also that  ``$Q$-MM-OFDM-IM $(Q, N,1)$'' and ``$Q$-MM-OFDM-IM $(Q, N, 1)$, QAM'' employ $Q$ disjoint constellations, each of which involves only one constellation point. This means that such schemes include only index symbols and the points in a $Q$-ary (PSK for the former and QAM for the latter) constellation form these symbols. 

\begin{figure}[t!]
		\centering
		\includegraphics[width=8cm,height=6cm]{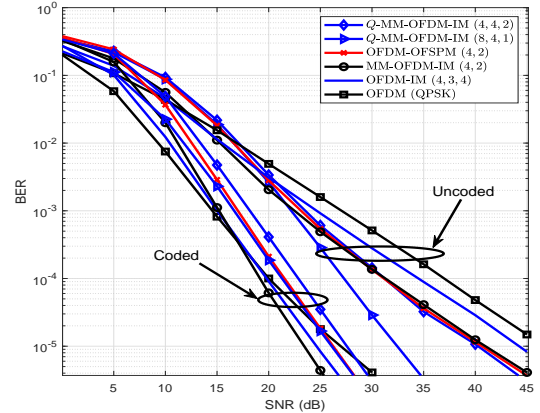}
		\caption{ Uncoded and coded BER comparison of $Q$-MM-OFDM-IM $(4,4, 2)$ and $Q$-MM-OFDM-IM $(8, 4, 1)$ with OFDM-OFSPM $(4, 2)$, MM-OFDM-IM $(4, 2)$, OFDM-IM $(4, 3, 4)$ and OFDM (QPSK).}
		\label{fig:fig1}
\end{figure}

Fig. \ref{fig:fig1} compares the  uncoded and coded BER performance of the proposed schemes\footnote{In Fig. \ref{fig:fig1}, BER curves of the uncoded schemes are obtained by employing the optimum ML detector, whereas the coded schemes perform hard-decision decoding based on the optimum ML detector.}, $Q$-MM-OFDM-IM $(4,4,2)$ and $Q$-MM-OFDM-IM $(8,4,1)$, with OFDM-OFSPM $(4,2)$, MM-OFDM-IM $(4,2)$, OFDM-IM $(4,3,4)$ and OFDM (QPSK).  SEs for uncoded $Q$-MM-OFDM-IM $(4,4,2)$; $Q$-MM-OFDM-IM $(8,4,1)$ and  OFDM-OFSPM $(4,2)$; MM-OFDM-IM $(4,2)$, OFDM-IM $(4,3,4)$ and OFDM (QPSK) are 2.5 bps, 2.25 bps and 2 bps, respectively. In the uncoded case, the $Q$-MM-OFDM-IM $(8,4,1)$ scheme is capable of achieving the same SE  as OFDM-OFSPM $(4, 2)$ by employing only index symbols, and it considerably outperforms the other schemes by introducing an additional diversity gain. Moreover, $Q$-MM-OFDM-IM $(4,4,2)$ outperforms OFDM-OFSPM $(4,2)$ and MM-OFDM-IM $(4,2)$ slightly, and OFDM-IM $(4,3,4)$ and OFDM (QPSK) considerably  at high SNR values although it achieves the highest SE.  In the coded case, the BER curves are shifted to the left. Hence, the proposed schemes are outperformed by the OFDM-IM and OFDM schemes for a larger BER range compared to the uncoded case. However, a similar behavior to the uncoded case is observed as the proposed schemes start to outperform conventional OFDM at relatively high SNR. Moreover, depending on the bit-to-index mapping and the constellation indices, the coded MM-OFDM-IM scheme outperforms the coded proposed schemes. 

\begin{figure}[t!]
		\centering
		\includegraphics[width=8cm,height=6cm]{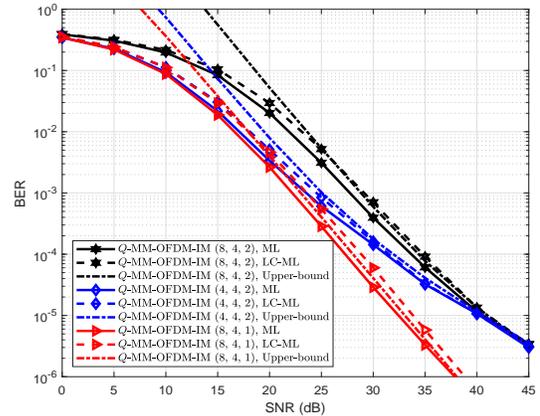}
		\caption{BER comparison of the proposed LC-ML detector with optimum ML detector for $Q$-MM-OFDM-IM $(8, 4, 2)$, $Q$-MM-OFDM-IM $(4, 4, 2)$ and $Q$-MM-OFDM-IM $(8, 4, 1)$.}
		\label{fig:fig2}
\end{figure}

Fig. \ref{fig:fig2} demonstrates the uncoded BER performance of the proposed LC-ML detector compared to the optimum ML detector based on \eqref{eq:eq2}  for the $Q$-MM-OFDM scheme when $Q\in\big\{4, 8\big\}$, $N=4$, and $M\in\big\{1, 2\big\}$. In this figure, ``$Q$-MM-OFDM-IM $(Q, N, M)$, LC-ML'' and ``$Q$-MM-OFDM-IM $(Q, N, M)$, ML'' stand for the $Q$-MM-OFDM-IM schemes applying the proposed LC-ML and optimum ML detectors, respectively. As seen from the figure, the performance of the proposed LC-ML detector is very close to that of the optimum ML detector, especially at low and high SNR values. Moreover, the performance loss in terms of SNR can be given approximately as 1.4 dB, 1 dB and 1.4 dB for $Q$-MM-OFDM-IM $(8, 4, 2)$, $Q$-MM-OFDM-IM $(4, 4, 2)$ and $Q$-MM-OFDM-IM $(8, 4, 1)$, respectively,  at a BER value of $10^{-3}$. The figure also shows the theoretical upper-bound results,``$Q$-MM-OFDM-IM $(Q, N, M)$, Upper-bound'', for the proposed scheme. As observed from the figure, upper-bound curves are consistent with computer simulations, especially at high SNR.   
        
\begin{figure}[t!]
		\centering
        \includegraphics[width=8cm,height=6cm]{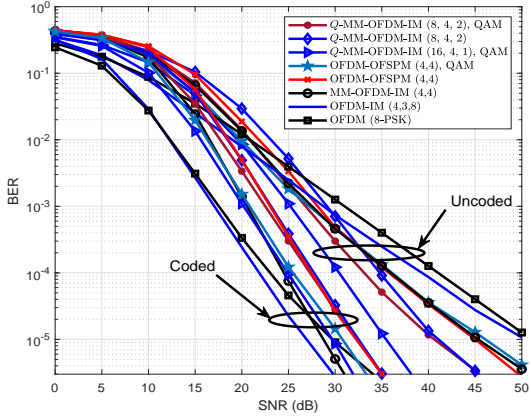}
		\caption{ Uncoded and coded BER comparison of $Q$-MM-OFDM-IM $(Q,N,M)$ and $Q$-MM-OFDM-IM $(Q,N,M)$, QAM with OFDM-OFSPM $(4,4)$, QAM, OFDM-OFSPM $(4,4)$,  MM-OFDM-IM $(4,4)$, OFDM-IM $(4,3,8)$ and OFDM (8-PSK) for $N=4, Q\in \big\{8, 16\big\}, M\in\big\{1, 2\big\}$.}
		\label{fig:fig3}
\end{figure}

In Fig. \ref{fig:fig3}, we compare the  uncoded and coded BER performance of the proposed schemes\footnote{ In Fig. \ref{fig:fig3}, the BER curves regarding the uncoded and coded $Q$-MM-OFDM-IM schemes are obtained by employing the proposed LC-ML detector and performing hard-decision decoding based on the proposed LC-ML detector, respectively. The remaining uncoded and coded schemes employ the optimum ML detector and perform hard-decision decoding based on the optimum ML detector at the receiver, respectively.}, $Q$-MM-OFDM-IM $(8,4,2)$, QAM, $Q$-MM-OFDM-IM $(8,4,2)$ and  $Q$-MM-OFDM-IM $(16,4,1)$, QAM, with OFDM-OFSPM $(4,4)$, QAM, OFDM-OFSPM $(4,4)$, MM-OFDM-IM $(4,4)$, OFDM-IM $(4,3,8)$ and OFDM (8-PSK) schemes. Here,  in the uncoded case, except for $Q$-MM-OFDM-IM $(16,4,1)$, the $Q$-MM-OFDM-IM and OFDM-OFSPM schemes exhibit the same SE of 3.25 bps whereas the remaining schemes, except for OFDM-IM, have the same SE of 3 bps. Also, the SE for OFDM-IM $(4,3,8)$ is 2.75 bps. As seen from the figure, $Q$-MM-OFDM-IM $(8,4,2)$, QAM, $Q$-MM-OFDM-IM $(8,4,2)$ achieve an outstanding BER performance at high SNR and outperform the OFDM-OFSPM $(4,4)$, QAM, OFDM-OFSPM $(4,4)$, and MM-OFDM-IM $(4,4)$ schemes as well as the OFDM-IM $(4, 3, 8)$, and OFDM (8-PSK) schemes by  providing almost 5 dB and 10 dB SNR gains, respectively, at a BER value of $10^{-5}$. These results arise from the fact that the $Q$-MM-OFDM-IM schemes achieve a similar SE as those of the other schemes by employing a lower order modulation. Such an order reduction results in higher minimum Euclidean distance for the modulation symbols as well as better error performance at high SNR values. Moreover, $Q$-MM-OFDM-IM $(16,4,1)$ achieves the best BER performance with an SNR gain of more than 10 dB at a BER value of $10^{-5}$ compared to the OFDM-OFSPM $(4,4)$, QAM, OFDM-OFSPM $(4,4)$, and MM-OFDM-IM $(4,4)$ schemes. Such a scheme achieves an additional diversity order due to the index symbols, whereas the diversity orders of the other schemes are limited by the minimum Hamming distance between the conventional modulation symbols.  In the coded case, OFDM-IM (4,3,8) achieves the best performance. However, at a relatively high SNR, $Q$-MM-OFDM-IM $(16, 4, 1)$ achieves a satisfactory performance with higher SE compared to the benchmarks.   

\section{Conclusion}\label{section:section6}
In this paper, we proposed a novel IM scheme that we call $Q$-MM-OFDM-IM. We showed that the proposed scheme encompasses conventional OFDM as a special case, and it is capable of providing the highest number of index symbols among other OFDM-IM schemes. We investigated the SE and BER performance of the proposed scheme and demonstrated that the proposed scheme can provide substantial performance improvement compared to MM-OFDM-IM, OFDM-OFSPM, OFDM-IM, and OFDM.               

As future work, the proposed scheme could be generalized by utilizing in-phase and quadrature dimensions of the modulation symbols as in \cite{Wen2017} and employing disjoint constellations with variable size as in \cite{Wen2018}.

\section*{acknowledgements}
The authors wish to acknowledge the support of the Bristol Innovation \& Research Laboratory of Toshiba Research Europe Ltd. They also would like to thank Dr. J. Vonk for his valuable  contributions to the discussions on the number of index symbols.

\bibliographystyle{IEEEtran}
\bibliography{main}

\end{document}